\documentclass[a4paper,conference]{IEEEtran}
\IEEEoverridecommandlockouts
\usepackage{cite}
\usepackage[utf8]{inputenc}
\usepackage{amsmath,amssymb,amsfonts}
\usepackage{algorithmic}
\usepackage{graphicx}
\usepackage{textcomp}
\usepackage{xcolor}
\usepackage{fancyhdr}

\def\BibTeX{{\rm B\kern-.05em{\sc i\kern-.025em b}\kern-.08em
    T\kern-.1667em\lower.7ex\hbox{E}\kern-.125emX}}
    
\usepackage{geometry}
\begin{document}

\title{Image Encryption Algorithm Using Natural Interval Extensions
{}
\thanks{ This work was supported by the funding groups CNPq/INERGE, Fapemig and Capes.}
}

\author{
\IEEEauthorblockN{Lucas Giovani Nardo}
\IEEEauthorblockA{\textit{Control and Modelling Group (GCOM)} \\
\textit{Department of Electrical Engineering}\\
\textit{Federal University of São João del-Rei}\\
São João del-Rei, Brazil \\
gnlucas@gmail.com}
\and
\IEEEauthorblockN{Arthur Mendes Lima}
\IEEEauthorblockA{\textit{Department of Electrical Engineering}\\
\textit{University of Brasília}\\
Brasília, Brazil \\
arthurlima67@yahoo.com.br}
\and
\IEEEauthorblockN{ Erivelton Geraldo Nepomuceno}
\IEEEauthorblockA{\textit{Control and Modelling Group (GCOM)} \\
\textit{Department of Electrical Engineering}\\
\textit{Federal University of São João del-Rei}\\
São João del-Rei, Brazil \\
nepomuceno@ufsj.edu.br}
\and
\IEEEauthorblockN{Janier Arias-Garcia}
\IEEEauthorblockA{\textit{Mechatronics, Control and Robotics (MACRO)} \\
\textit{Department of Electronic Engineering}\\
\textit{Federal University of Minas Gerais}\\
Belo Horizonte, Brazil \\
janier.arias@gmail.com}
}

\maketitle

\begin{abstract}
It is known that chaotic systems have widely been used in cryptography. Generally, floating point simulations are used to generate pseudo-random sequence of numbers. Although, it is possible to find some works on the degradation of chaotic systems due to finite precision of digital computers, little attention has been paid to exploit this limitation to formulate efficient process for image encode. This article proposes a novel image encryption method using natural interval extensions. The sequence of arithmetic operations is different in each natural interval extension. This is what we need to produce two different sequences; the difference between these sequences is used to generate the lower bound error, which has been shown to present satisfactory pseudo-random properties.  The approach has been successfully tested using the Chua’s circuit as the chaotic system. The secret key has presented good  properties for encrypting the Lena image.
\end{abstract}


\section{Introduction}

It has been an integral part of human nature to maintain control of the access to information. For this reason, encryption has received such attention over the past few years. For example, encryption takes place in bank and cryptocoins transactions \cite{carrott2017secure}.  Additionally, this area has such importance in image encryption \cite{chai2017image}. In Computer Science and Electrical Engineering fields, more robust and effective encryption methods have emerged as demonstrated by the linear congruential method  \cite{jain1990art}, by the use of irrational numbers \cite{liu2017information} and also for chaotic systems \cite{lv2015perturbation}.

Chaotic dynamical systems present interesting properties as its transitivity, the high density of the periodic points of the function $f$ in metric space and its sensitive to initial conditions \cite{zhang2005image},\cite{banks1992devaney}. Therefore, these systems can generate pseudo-randomness sequences, which can be used in cryptography. In fact, Herring e Palmore \cite{HP1989} have already told that pseudo-random number generators  are examples of deterministic chaotic dynamical systems.

In the information security area, chaotic systems have been vastly studied and several methods have been emerged.  Fridrich \cite{fridrich1997image} used only the chaotic properties of the baker map;  Ismail et al. \cite{ismail2015generalized} added two parameters to the classical fractional logistic equation to improve its flexibility and control; and Zhang \cite{zhang2016image} used the hyper-chaotic Chen’s system, diffusion and shuffling operations to encrypt images. One of the key problems faced by researchers in this area is the degradation of chaotic systems due to finite precision of digital computers, as reported by  Li et al. \cite{li2005dynamical}. The last few years many attempts have been investigated to overcome this problem, such as the use of high finite precision, cascading multiple chaos systems, switching multiple chaos systems, coupling different chaotic systems or pseudo randomly perturbing the chaotic system.  The reader is invited to read the work by Cao et al. \cite{lv2015perturbation} for more information on these methods. 

Although, many researchers have succeed to reduce the degradation of the chaotic properties of digital systems, little attention has been paid to exploit this limitation to formulate encryption algorithms. Instead of seeing the finite digital precision as a problem, this paper proposes a novel image encryption method using natural interval extensions. When solving a chaotic system by means of numerical computation, it is verified that, based on two natural interval  extensions (for more details, see section \ref{lowerbounderror}), starting from the same set of parameters and initial conditions, after a certain number of iterations, the results of each simulation of the system diverge. Such an event could not occur due to the exactly equal initial sets of parameters and conditions for each simulation. This happens because of the constructive limitations of computers and the IEEE 754-2008 floating point standard \cite{zuras2008ieee}. As observed in \cite{nepomuceno2016lower,nepomuceno2017lower}, the sequence of arithmetic operations is different in each natural interval extension. This is what we need to produced two different sequences to generate the lower bound error, which has been shown to present satisfactory pseudo-random properties.

\begin{figure*}[!ht]
	\centering
	\includegraphics[width=0.7\linewidth]{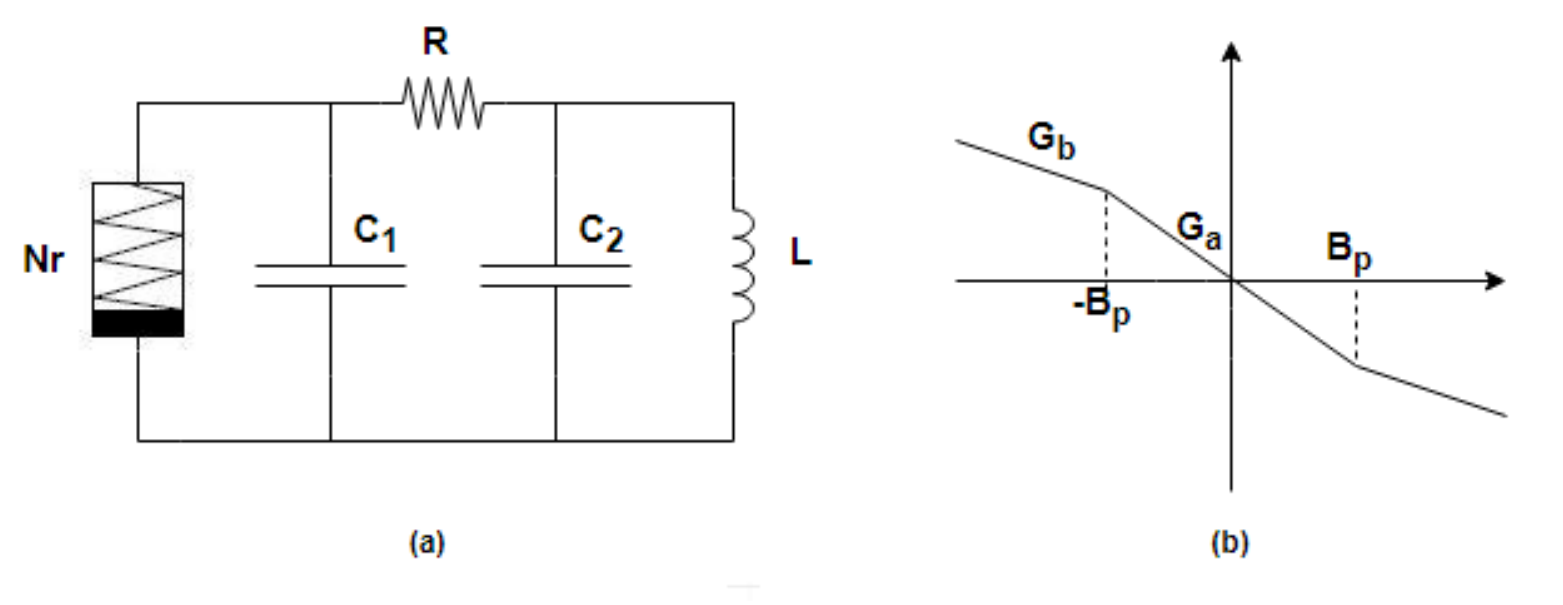}
	\caption{(a) Chua's circuit. (b) Chua's diode curve. $G_a$, $G_b$ and $B_p$ are the slopes and the breaking points of the nonlinear element, respectively.}
	\label{chua}
\end{figure*}

\begin{figure*}[!t]
    \centering
    \begin{equation}
\left\{ \begin{array}{rcl}
C_1 \cfrac{dv_{c_1}}{dt} &=&\cfrac{v_{c_2} - v_{c_1}}{R} - i_R(v_{c_1})\\ C_2 \cfrac{dv_{c_2}}{dt} &=&\cfrac{v_{c_1} - v_{c_2}}{R} + i_L\\
L \cfrac{di_L}{dt} &=& - v_{c_2} 
\end{array}\right.
\label{eq.Chua}
\end{equation}
\end{figure*}

\begin{figure*}[!t]
    \centering
\begin{equation}
i_R(v_{C_1})=\left\{ \begin{array}{rcl}
{G_b v_{C_1} + B_p(G_b - G_a)}, &\mbox{if}
&v_{C_1}< -B_p \\ {G_a v_{C_1} }, &\mbox{if}
&|v_{C_1}| \leq B_p \\
{G_b v_{C_1} + B_p(G_a - G_b)}, &\mbox{if}
&v_{C_1})> B_p 
\end{array}\right.
\label{eq.DiodoChua}
\end{equation}
\end{figure*}

This paper is developed as follows: a brief introduction, followed by a bibliographic review was presented in this section. Section II points out important concepts for understanding the rest of the text. The methods used in this work, as well as the results, are  shown in Sections III and IV, respectively. Finally, section  V contains the conclusion of the paper.

\section{Preliminary Concepts}

\subsection{Chua's circuit}

The circuit developed by Chua et al. \cite{chua1993universal}, exhibits nonlinear behaviour, such as a spiral attractor and double-scroll attractor. Since then, this circuit (Fig. \ref{chua}) has been extensively studied and simulated computationally. The circuit is composed by linear passive elements: two capacitors, an inductor and a resistor, which are connected to an active, nonlinear element called the Chua's diode (Nr), as shown in Fig. \ref{chua}a. Therefore, according to Kirchhoff's law, it is possible to obtain the differential equations which represents the circuit's dynamic, \eqref{eq.Chua}. The resistive effect of the inductor is considered as imperceptible. The current through the nonlinear element, $ i_R (v_{C_1}) $ is given by \eqref{eq.DiodoChua}. Fig. \ref{chua}b displays the nonlinear behaviour of the Chua's diode, given by the relation \textit{voltage} x \textit{current} of the component.

\subsection{Lyapunov Exponent}
There are many definitions about chaos. However, the concepts of Lyapunov exponents are the most influential work present in literature. The Lyapunov Exponent (LE), is a method which quantifies the exponential divergence of initially close orbits. The presence of a positive LE indicates chaos. In literature, there are numerous methods to determine the LE, as the Kantz's method \cite{kantz1994robust}, for example. In Kantz's approach, he considers the following equation:

\begin{equation}
\begin{aligned}
 S(\Delta n) = \dfrac{1}{N-m}\sum\limits_{n=m+1}^N \\
\times ln \left(\dfrac{1}{|\eta_n|}\sum\limits_{x_{n'}\epsilon\eta_n}|x_{n'+\Delta n}-x_{n+\Delta n}|\right)
\label{eq.lyapdes}
\end{aligned}
\end{equation}

\noindent where $\eta_n$ is the set of all others delay vectors in an
$\epsilon$-neighborhood of the vector $x_n$ (data from trajectories
of the system under investigation) and $|\eta_n|$ is
the number of elements in $\eta_n$. The Lyapunov exponent
can be estimated by searching for a linear scaling
in plot $S(\Delta n)$ versus $\Delta n$ \cite{nepomuceno2016very}.

By the characteristics mentioned above, it is possible to relate the importance of this topic in cryptography. Since the system (Chua's circuit) used to encrypt the image is chaotic, the process of decryption, without knowing the seed and the secret key is computationally expensive, making it a difficult task.

\subsection{The lower bound error}
\label{lowerbounderror}
The lower bound error is used to analyze the error propagation in numerical simulations \cite{nepomuceno2017lower}. For the understanding of this tool,  orbits, pseudo-orbits and natural interval extensions are defined in this section.

\textbf{Definition 1}: an orbit is a sequence of values of a map or system, represented by $x_i = [x_0;x_1;x_2;x_3...x_i]$.

Results of numerical simulation, due to truncation and rounding errors, inherent of a computer, cannot fit into a true orbit, therefore they are called pseudo-orbits.


\textbf{Definition 2}: a pseudo-orbit is an approximation of the true orbit, represented by ${\hat{x_i}}=[\hat{x_0},\hat{x_1},\hat{x_2},\hat{x_3}...\hat{x_i}]$ which accepts the relation $|x_i-\hat{x}_{i}|\le\delta$, where $\delta$ is the associated error.

As described by Nepomuceno and Martins \cite{nepomuceno2016lower}, a natural interval extension is defined: 

\textbf{Definition 3}: a natural interval extension of a function $f$ is an interval-valued function $F$ of an interval variable $X$, with the property $F(x) = f(x)$, where by an interval it is meant to be a closed set of real numbers $x \in \Re $ such that $X=[\underline{X},\overline{X}]={x:\underline{X} \leq x \leq \overline{X}}$.

Using the Chua's circuit equations, examples of natural interval extensions are shown by (\ref{eq:2}) and (\ref{eq:3}).

\begin{equation}
\label{eq:2}
C_1\dfrac{dv_{C_1}}{dt} = \dfrac{v_{C_2}-v_{C_1}}{R}-i_R(v_{c_1}) 
\end{equation}

\begin{equation}
\label{eq:3}
C_1\dfrac{dv_{C_1}}{dt} = \dfrac{v_{C_2}}{R}-\dfrac{v_{C_1}}{R}-i_R(v_{c_1})
\end{equation}

The lower bound error is established by the following definition:

\textbf{Definition 4}: given two pseudo-orbits $\hat{x}_{a,n}$ and $\hat{x}_{b,n}$, arising from two different natural interval extensions of the function $f(x)$, the lower bound error $\delta$ is given by:

\begin{equation}
\label{eq:2a}
\delta=\dfrac{|\hat{x}_{a,n}-\hat{x}_{b,n}|}{2}.
\end{equation}

\subsection{Cryptography}

Cryptography is the science which studies techniques to make data illegible. In this way, it is possible to transmit all information securely. To perform the decryption, one should be aware of  cryptographic key \cite{mollin2000introduction}. 

The encryption and decryption can be done using the bit-wise XOR operation, because the probability of the XOR output being zero or one is $50\%$ and by the following propriety: $(A \oplus B) \oplus B = A \oplus 0 = A $. In other words, using the cryptographic key B twice in the document that you want to encrypt A, the result remains A. This property represents the entire cryptographic process.   

To ensure that the encryption process is good enough to make the image illegible, there are several ways to testify this: the correlation coefficient of adjacent pixels randomness test, the Shannon entropy test and the distribution of pixel in an image plotting an histogram.
In a histogram, when the encryption process is performed, the cipher image must be uniform, in other words, the frequencies of the pixels must be approximately equal to all color intensities. So doing, the cipher image does not bring any relevant information.

It is well known that in plain-images, the adjacent pixels are strongly correlated with each other. Therefore, in a cipher image, the correlation coefficient in horizontal, vertical and diagonal directions are expected to be close to zero.  The correlation coefficient of adjacent pixels randomness test measures this correlation by (7) \cite{dianocu2017correlation}.

\begin{equation}
\label{eq:7}
\rho(X,Y)=\dfrac{E[(X-\mu_X)(Y-\mu_Y)]}{\sigma_X\sigma_Y}
\end{equation}

\noindent where $X$ represents the series of pixels at position, $Y$ represents the series of adjacent pixels, $\mu$ and $\sigma$ are the mean and the standard deviation values, respectively, and $E$ is the mathematical expectation.

The Shannon entropy is a tool to measure the randomness in a communication systems. It is defined by (8) \cite{luo}:

\begin{equation}
\label{eq:7a}
H(X)=\sum\limits_{i=1}^{2^N-1}P_ilog_2 \dfrac{1}{P_i}
\end{equation}

\noindent where $H(X)$ is the entropy (bits), $X$ is a symbol and $P_i$ is the probability value of symbol $X$.

In image encryption, there are 256 values that each pixel can be defined. Therefore, for a cipher image, the expected value is $H(X)=8$ bits. 

\section{Methodology}

A fundamental part of the method is the cryptographic key, which is the pseudo-randomness sequence. The sequence is generated simulating the Chua's circuit using the fourth order Runge-Kutta method, an integration step equal to $10^{-6}$ and the most important, the two natural interval extensions presented by \eqref{eq:2} and \eqref{eq:3} (see section \ref{lowerbounderror}). This system has been chosen to apply the method, because it is a benchmark in the study of dynamical systems and the most important, its chaotic properties. The following Chua's parameters was used to generate the two pseudo-orbits: $C_1=10nF$, $C_2=100nF$,  $L=19mH$, $R=1.8k\Omega$, $G_a=-0.68mS$, $G_b=-0.37mS$, $Bp=1.1V$, $V_{C1}=-0.5V$, $V_{C2}=-0.2V$, $I_{L}=0A$. Afterwards, to encrypt an image, we have used the following steps, adapted from \cite{lv2015perturbation}:

\begin{itemize}
    \item \textbf{Step 1}: For an image with $M \times N$ pixels, perform two simulations with different
    natural interval extensions of the Chua's circuit with $2000 + M \times N - 1$ iterations. The first 2000 points generated will be discarded. This is due to the fact that at the beginning of the simulation, the two pseudo-orbits are close to each other, making the generated sequence easy to identify. We have chosen 2000 points according to the critical time simulation described in \cite{NM2017}.   
    \item \textbf{Step 2}: After the two sequences $S_1$ and $S_2$ generated, the logarithm of the lower bound error is done, generating a single sequence S:
    \begin{equation}
            S=log_{10}\dfrac{|S_1-S_2|}{2}.
    \end{equation}
     \item \textbf{Step 3}: The normalizing process of the sequence S is done as follows:
     \begin{equation}
         S_n = uint8(mod(S\times10^{15},256)),
     \end{equation}
     
    which $S_n$ is the normalized sequence. Uint8 is an algorithm available on the latest release of the software \textit{Matlab}, which converts the sequence into 8-bit positive integer and mod represents the modulo operation: the rest of the division $S \times 10^{15}$ by $256$. This is necessary as tested images are 8-bit gray using a  pixel matrix with number between 0 (black tone) and 255 (white tone).
    \item \textbf{Step 4}: With the key $S_n$ and the media to be encrypted in the same numeric format, the bit-wise XOR operation is performed.
\end{itemize}

The standard gray $256 \times 256$ Lena image was used to assess the algorithm performance. All data, routines and simulations used in this work were generated using the Matlab software and are available upon request.



\begin{figure*}[!ht]
	\centering
	\includegraphics[width=0.7\linewidth]{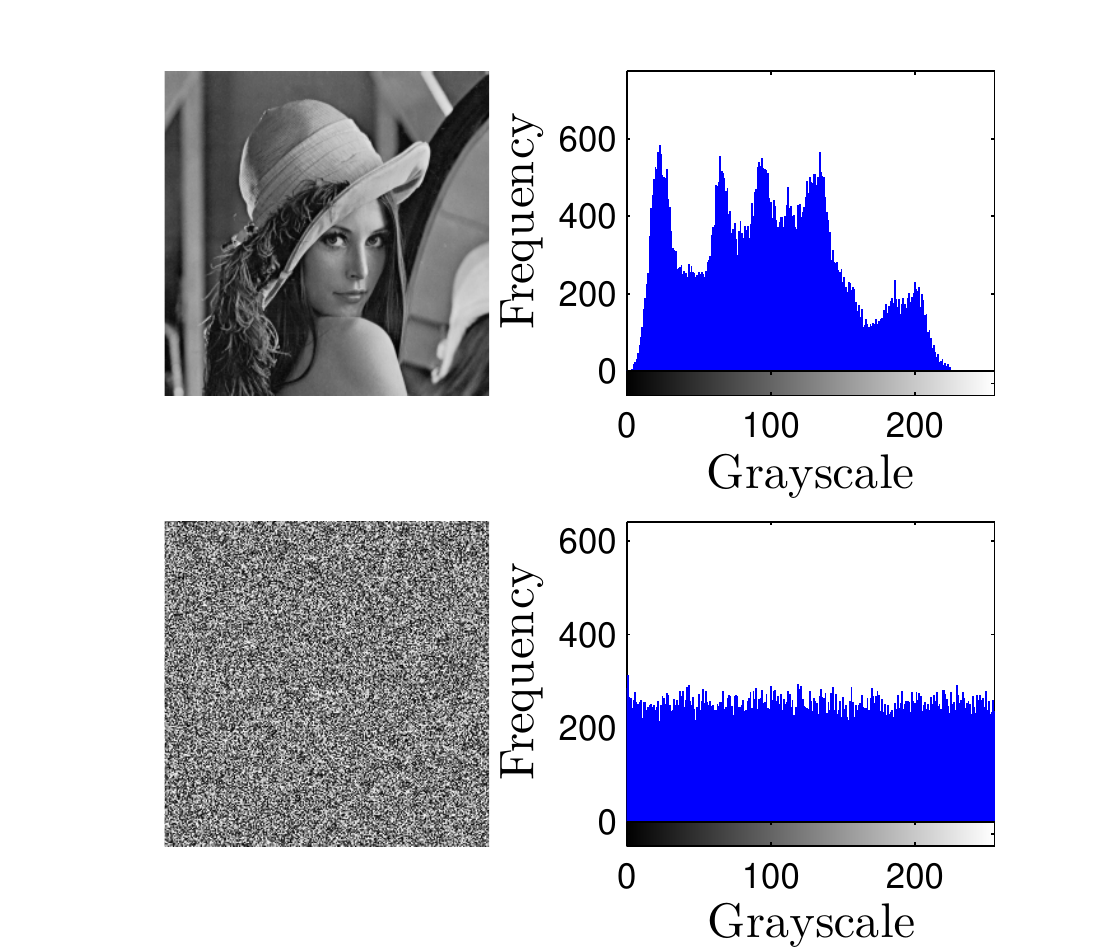}
	\caption{The plain-image Lena as well as the cipher image herewith their histograms.}
	\label{lena2}
\end{figure*}

\begin{figure}[ht!]
	\centering
	\includegraphics[width=1\linewidth]{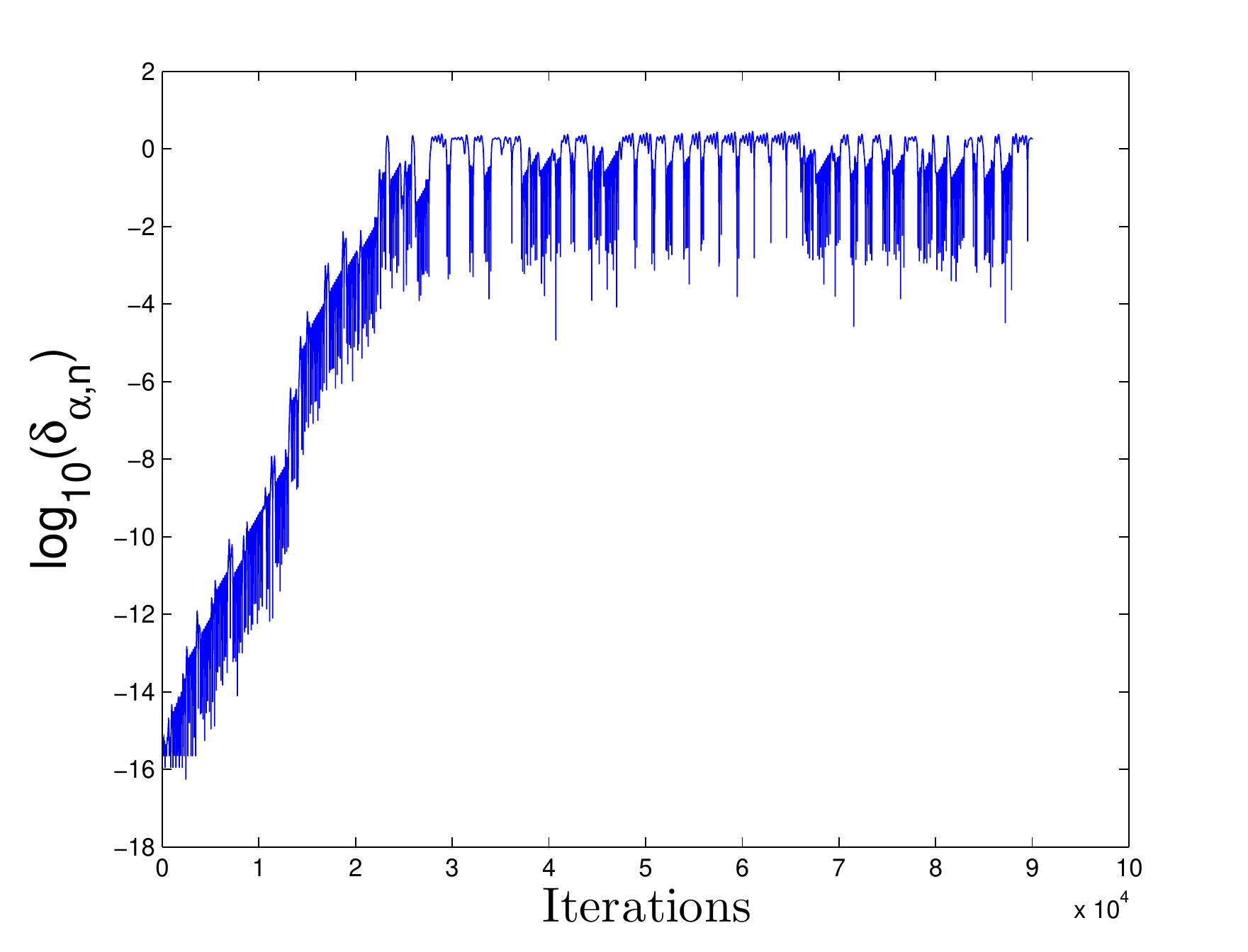}
	\caption{The lower bound error obtained from two interval extensions.}
	\label{crip2}
\end{figure}

\section{Results and Discussion}

The lower bound error, which is obtained from the parameters specified in the previous section, is shown in the Fig. \ref{crip2}. The two natural interval extensions used are \hspace{1cm} $C_1\dfrac{dv_{C_1}}{dt} = \dfrac{v_{C_2}-v_{C_1}}{R}-i_R(v_{c_1})$ and $C_1\dfrac{dv_{C_1}}{dt} = \dfrac{v_{C_2}}{R}-\dfrac{v_{C_1}}{R}-i_R(v_{c_1})$.  The LE ($\lambda$) was calculated by the Kantz's method, which resulted in a positive value equal to $\lambda=0.199$, demonstrating its chaotic behavior and pseudo-randomness.

\begin{figure*}[ht!]
	\centering
	\includegraphics[width=1.0\linewidth]{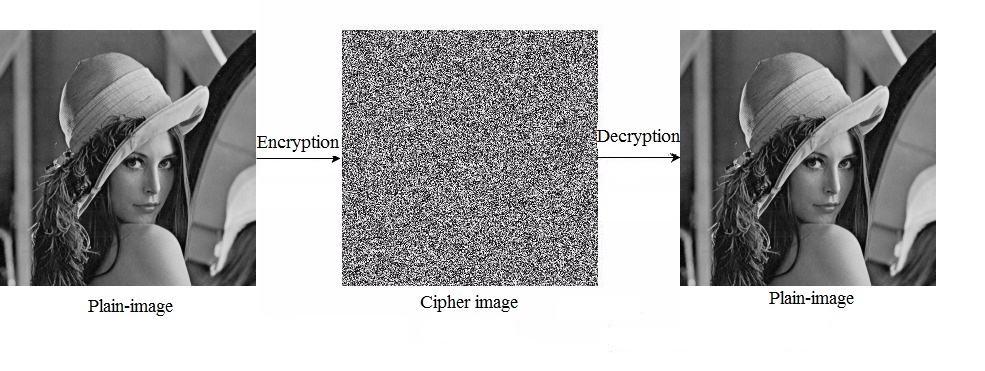}
	\caption{By performing the bit-xor operation twice, the encryption and decryption process of an image is made, which represents the entire cryptographic process.}
	\label{lena2a}
\end{figure*}

The Lena image was encrypted, after all the steps performed. Fig. \ref{lena2} shows the plain-image and the cipher image along with their respective histograms. The original Lena image features a non-uniform distribution in the graphic. However, when it is encrypted, the histogram features a uniform distribution, which each color intensity level has the same frequency, approximately, becoming an illegible image. It is worth noting that with the decryption process, the image becomes legible again, as shown in Fig. \ref{lena2a}.

The entropy test and correlation between two adjacent pixels was executed and the results were slightly close to the expected value (see Table 1).

\begin{table}[ht!]
\centering
\caption{Tests for the cipher image}
\label{my-label}
\begin{tabular}{c|c|c|c|c}
\hline
\multicolumn{3}{c|}{Correlation Coefficient} &
Entropy & References \\ \cline{1-3}
Horizontal & Vertical & Diagonal &  &  \\ \hline
0.0028 & 0.0059 & 0.0031 & 7.9969 & This work \\ \hline
0.0016 & 0.0025 & 0.0003 & 7.9998 & C. Li et al.\cite{tent} \\ \hline
0.00083 & 0.00223 & 0.00650 & 7.9826 & Y. Luo et. al \cite{luo} \\ \hline
\end{tabular}
\end{table}

\section{Conclusion}

In this paper, a novel image encryption method has been presented. This method is based on the concept of natural interval extension and the fact of limitation of numerical representation presented on computers. The proposed method was very  efficient, producing a pseudo-random sequence with good cryptographic properties and encrypting the Lena image. The entropy measure and the correlation coefficients calculated using our approach has been shown to be as efficient as other works presented in literature. Furthermore, it is important to emphasise that the method evidence the random characteristic of the error.

As a future work, the authors propose the study and the accomplishment of other tests, with the intention of analysing the computational performance and improving the proposed method, also constructing an embedded system for the use of cryptography in the most diverse fields.

\section*{Acknowledgment}

The authors thank the anonymous reviewers for their constructive comments which helped to improve the manuscript. 


\end{document}